\documentstyle[11pt]{article}

\def \lket {|}
\def \rket {\rangle}

\newcommand{\ket}[1]{\lket #1\rket}

\newtheorem{Definition}{Definition}
\newtheorem{Theorem}{Theorem}
\newtheorem{Lemma}{Lemma}

\newcommand{\proof}{\noindent {\bf Proof: }}
\newcommand{\qed}{$\Box$}
\newcommand{\comment}[1]{}

\begin{document}
\title{Polynomial Degree and Lower Bounds in Quantum Complexity:
       Collision and Element Distinctness with Small Range}
\author{
Andris Ambainis\thanks{
Department of Combinatorics and Optimization,
Faculty of Mathematics, 
200 University Avenue West,
Waterloo, ON  N2L 3G1, Canada
e-mail:{\tt ambainis@math.uwaterloo.ca}. 
Supported by IQC University Professorship and CIAR.
This work done while at 
University of Latvia.}}

\date{}

\maketitle

\begin{abstract}
We give a general method for proving quantum lower bounds 
for problems with small range.
Namely, we show that, for any symmetric problem defined on 
functions $f:\{1, \ldots, N\}\rightarrow\{1, \ldots, M\}$,
its polynomial degree is the same for all $M\geq N$.
Therefore, if we have a quantum query
lower bound for some (possibly, quite large) 
range $M$ which is shown using the polynomials method, we immediately get the
same lower bound for all ranges $M\geq N$.
In particular, we get $\Omega(N^{1/3})$ and 
$\Omega(N^{2/3})$ quantum lower bounds 
for collision and element distinctness with small range.
As a corollary, we obtain a better lower bound on the
polynomial degree of the two-level AND--OR tree.
\end{abstract}

\date{}

\maketitle

\section{Introduction}

Quantum computing provides speedups for many search problems.
The most famous example is Grover's algorithm \cite{Grover}, which computes OR of 
$N$ variables with $O(\sqrt{N})$ queries.
Other examples include counting \cite{Counting}, 
estimating mean and median \cite{Grover97,NW}, 
finding collisions \cite{BCollision}, 
determining element distinctness \cite{BDistinct,Ambainis},
finding triangles in a graph \cite{MSS} and
verifying matrix products \cite{BS}.
For many of these problems, we can also prove
that known quantum algorithms are optimal or nearly optimal.

In at least two cases, the lower bounds match the best known algorithm
only with an additional ``large range" assumption. For example, consider
the collision problem \cite{BCollision,AS} which models
collision-free hash functions. We have to
distinguish if a function $f:\{1, \ldots, N\}\rightarrow\{1, \ldots, M\}$
is one-to-one or two-to-one. A quantum algorithm can solve the problem
with $O(N^{1/3})$ queries (evaluations of $f$) \cite{BCollision}, which is better than 
the $\Theta(N^{1/2})$ queries required classically. A lower bound by Aaronson and Shi \cite{AS}
says that $\Omega(N^{1/3})$ quantum queries are required if $M\geq 3N/2$.
If $M=N$, the lower bound becomes $\Omega(N^{1/4})$. 

A similar problem exists for element distinctness. 
(Again, we are given $f:\{1, \ldots, N\}\rightarrow\{1, \ldots, M\}$
but $f$ can be arbitrary and 
we have to determine if there are $i, j$, $i\neq j$, $f(i)=f(j)$.)
If $M=\Omega(N^2)$, the lower bound is $\Omega(N^{2/3})$ \cite{AS}, 
which matches the best algorithm \cite{Ambainis}.
But, if $M=N$, the lower bound is only $\Omega(\sqrt{N})$ or $\Omega(\sqrt{N\log N})$, 
depending on the model \cite{BDistinct,HNS}.  

Thus, it might be possible that a quantum algorithm could
use the small $M$ to decrease the number of queries. 
While unlikely, this cannot be ruled out.
Remember that classically, sorting requires $\Omega(N \log_2 N)$
steps in the general case but only $O(N)$ steps if the 
items to be sorted are all from the set $\{1, \ldots, N\}$
(Bucket Sort, \cite{CLR}).

In this paper, we show that the collision and element distinctness problems 
require $\Omega(N^{2/3})$ and $\Omega(N^{1/3})$ queries even if the
range $M$ is equal to $N$. Our result follows from a general
result on the polynomial degree of Boolean functions.

We show that, for any symmetric
property $\phi$ of functions $f:\{1, 2, \ldots, N\}\rightarrow \{1, 2, \ldots, M\}$, 
its polynomial degree is the same for all $M\geq N$.
The polynomial degree of $\phi$ provides a lower bound for both classical  
and quantum query complexity. (This was first shown by Nisan and Szegedy \cite{NS}
in the classical case and then extended to the quantum case by Beals et al. \cite{Beals}
for $M=2$ and Aaronson \cite{Aaronson,AS} for $M>2$.) 
Thus, one can prove lower bounds on quantum query complexity of a function $\phi$
by lower-bounding the polynomial degree of $\phi$.
This is known as the polynomials method for proving quantum lower
bounds \cite{Beals,BWSurvey,AS}.

Our result means that, if we have a quantum lower bound for a symmetric property $\phi$ shown 
by the polynomials method for some range size $M$, we also have the same quantum lower bound for all $M\geq N$. 
As particular cases, we get lower bounds on the collision and element distinctness problems
with small range. Since many quantum lower bounds are shown using
the polynomial degree method, our result may have other 
applications.

A corollary of our lower bound on element distinctness with small
range is that the polynomial degree of the two level AND--OR tree
on $N^2$ variables is $\Omega(N^{2/3})$.
This improves over the previously known lower bound 
of $\Omega(\sqrt{N \log N})$ by Shi \cite{Shi1}.

{\bf Related work.}
The $\Omega(N^{1/3})$ lower bound for the collision problem 
with small range was independently discovered by the author of this paper 
and Kutin \cite{Kutin}, at about the same time, with completely
different proofs.
Kutin \cite{Kutin} takes the proof of $\Omega(N^{1/3})$ 
lower bound for the collision problem with a
large range \cite{AS} and changes it so that it works for all $M\geq N$.
Our result is more general because it applies to any symmetric property and 
any lower bound shown by the polynomials method.
On the other hand, Kutin's proof has the advantage that it
also simplifies the lower bound for the collision problem with large range
by Aaronson and Shi \cite{AS}.

\section{Preliminaries}

\subsection{Quantum query model}

Let $[k]$ denote the set $\{1, \ldots, k\}$.
Let $f$ be a function from $[N]$ to $[M]$.
Let ${\cal F}(N, M)$ be the set of all $f:[N]\rightarrow[M]$.
We are given a function $f\in {\cal F}(N, M)$ by an oracle
that answers queries. In one query, we can give $i$ to the oracle
and it returns $f(i)$ to us. 
 
We would like to know whether $f$ has a certain property
(for example, whether $f$ is one-to-one). More formally, we would like to 
compute a partial function $\phi:{\cal F'}\rightarrow \{0, 1\}$, where
${\cal F'}\subseteq {\cal F}(N, M)$.
In particular, we are interested in the following two properties:

{\bf Problem 1: Collision.} 
$\phi(f)=1$ if the input function $f$ is one-to-one. 
$\phi(f)=0$ if $f$ is two-to-one (i.e., if, for every $k\in[M]$,
there are either zero or two $x\in [N]$ satisfying $f(x)=k$).
$\phi(f)$ is undefined for all other $f$.

{\bf Problem 2: Element distinctness.} 
$\phi(f)=1$ if the input function $f$ is one-to-one. 
$\phi(f)=0$ if there exist $i, j$, $i\neq j$, $f(i)=f(j)$.

A quantum algorithm with $T$ queries
is just a sequence of unitary transformations
\[ U_0\rightarrow O_f\rightarrow U_1\rightarrow O_f\rightarrow\cdots
\rightarrow U_{T-1}\rightarrow O_f\rightarrow U_T.\]
The $U_j$'s can be arbitrary unitary transformations that do not depend
on $f(1), \ldots, f(N)$. $O$ is a query (oracle) transformation.
To define $O_f$, we represent basis states as $|i, b, z\rangle$ where
$i$ consists of $\lceil \log N\rceil$ bits, 
$b$ is $\lceil\log M\rceil$ bits and
$z$ consists of all other bits. Then, $O_f$ maps
$\ket{i, b, z}$ to $\ket{i, (b+f(i))\bmod M, z}$. 

The computation starts with a state $|0\rangle$.
Then, we apply $U_0$, $O_f$, $\ldots$, $O_f$,
$U_T$ and measure the final state.
The result of the computation is the rightmost bit of
the state obtained by the measurement.

The quantum algorithm computes
$\phi$ with error $\epsilon$ if, 
for every $f$ such that $\phi(f)$ is defined,
the probability that the rightmost bit 
of $U_T O_f U_{T-1} \cdots O_f U_0\ket{0}$ 
equals $\phi(f)$ is at least $1-\epsilon$.
(Throughout this paper, $\epsilon$ is an arbitrary but fixed
value, with $0<\epsilon<1/2$.)

\subsection{Polynomial lower bound}
\label{sub:polylb}

We can describe a function $f:[N]\rightarrow[M]$ by $N\times M$ Boolean 
variables $y_{ij}$ which are 1 if $f(i)=j$ and 0 otherwise.
Let $y=(y_{11}, \ldots, y_{NM})$.
\comment{Then, we have

\begin{Lemma}
\label{ALemma}
\cite{Aaronson,AS}
The state of a quantum algorithm after $T$ queries is
\[ \ket{\psi}=\sum_{i=1}^D a_i(y_{11}, \ldots, y_{NM}) \ket{i}, \]
with $a_i(y_{11}, \ldots, y_{NM})$ being polynomials in 
$y_{11}$, $\ldots$, $y_{NM}$ of degree at most $T$.
\end{Lemma}}

\begin{Definition}
\label{def1}
We say that a polynomial $P$ $\epsilon$-approximates the property
$\phi$ if
\begin{enumerate}
\item
$\phi(f)=1$ implies $1-\epsilon\leq P(y)\leq 1$
for $y=(y_{11}, \ldots, y_{NM})$ corresponding to $f$;
\item
$\phi(f)=0$ implies $0\leq P(y)\leq \epsilon$
for $y=(y_{11}, \ldots, y_{NM})$ corresponding to $f$;
\item
If $\phi(f)$ is undefined, then
$0\leq P(y)\leq 1$ for the corresponding $y$.
\end{enumerate}
A polynomial $P$ approximates $f$ if it $\epsilon$-approximates $f$
for some fixed $\epsilon<1/2$
\end{Definition}

$P$ is allowed to take any value if $y$ does not
correspond to any $f$. (This happens if for some
$i\in[N]$ there is no or there is more than one 
$j\in[M]$ with $y_{ij}=1$.)

\comment{Lemma \ref{ALemma} implies  }

\begin{Lemma}
\cite{Aaronson,AS}
If a quantum algorithm computes $\phi$ with error $\epsilon$ using $T$ queries
then there is a polynomial $P(y_{11}, \ldots, y_{NM})$ of degree at most
$2T$ that $\epsilon$-approximates $\phi$.
\end{Lemma}

A lower bound on the number of queries can be then shown by
proving that such a polynomial $P$ does not exist.
For the collision and element distinctness problems, we have
\begin{Theorem}
\label{STheorem}
\cite{Shi,AS}
\footnote{More 
precisely, Shi \cite{Shi,AS} proved that any polynomial 
approximating another problem, the {\em half two-to-one} problem, 
has degree $\Omega(N^{1/3})$. 
He then used that to deduce that $\Omega(N^{1/3})$
and $\Omega(N^{2/3})$ quantum queries are needed for 
the collision problem (when $M\geq \frac{3N}{2}$)
and the element distinctness problem (when $M=\Omega(N^2)$). 
His proof can be easily modified
to show a lower bound on the degree of polynomials 
approximating the collision and element distinctness problems.}
\begin{enumerate}
\item
If a polynomial $P$ approximates the collision property for $M\geq \frac{3N}{2}$,
the degree of $P$ is $\Omega(N^{1/3})$;
\item
If a polynomial $P$ approximates the element distinctness property for $M=\Omega(N^2)$,
the degree of $P$ is $\Omega(N^{2/3})$;
\end{enumerate}
\end{Theorem}

Therefore, $\Omega(N^{1/3})$
and $\Omega(N^{2/3})$ queries are required to
solve the collision problem and element distinctness problem
if the range $M$ is sufficiently large.
Previously, only weaker lower bounds of $\Omega(N^{1/4})$ \cite{AS} and
$\Omega(\sqrt{N\log N})$ \cite{HNS} were known if $M=N$.

\section{Results}

We call a property $\phi$ {\em symmetric} if,
for any $\pi\in S_N$ and $\sigma\in S_M$,
\[ \phi(f) = \phi(\sigma f \pi).\]
That is, $\phi(f)$ should remain the same if we permute the input
set $\{1, \ldots, N\}$ before applying $f$ or permute the output set
$\{1, \ldots, M\}$ after applying $f$.
The collision and element distinctness properties are both symmetric.

Our main result is

\begin{Theorem}
\label{MainThm}
Let $\phi:{\cal F}'\rightarrow \{0,1\}$, 
${\cal F}'\subseteq {\cal F}(N, M)$ be symmetric.
Let $\phi'$ be the restriction of $\phi$ to 
$f:[N]\rightarrow [N]$.
Then, the minimum degree of a polynomial (in $y_{ij}$,
$i\in[N]$, $j\in[M]$)
approximating $\phi$ is equal to 
the minimum degree of a polynomial (in $y_{ij}$,
$i\in[N]$, $j\in[N]$)
approximating $\phi'$.
\end{Theorem}

Theorems \ref{STheorem} and \ref{MainThm} 
imply that $\Omega(N^{1/3})$
and $\Omega(N^{2/3})$ queries are needed to
solve the collision and element distinctness problems,
even if $M=N$. (For $M<N$, these problems do
not make sense because they both involve $f$ being one-to-one
as one of the cases.)

The proof of Theorem \ref{MainThm} is in two steps.
\begin{enumerate}
\item 
We describe a different way to describe an input function $f$ by
variables $z_1$, $\ldots$, $z_M$ instead of $y_{11}, \ldots, y_{NM}$.
We prove that a polynomial of degree $k$ in $z_1$, $\ldots$, $z_M$ 
exists if and only if a polynomial of degree $k$ in $y_{11}$, $\ldots$, $y_{NM}$ exists.
\item
We show that a polynomial $Q(z_1, \ldots, z_M)$ for $M>N$ exists
if and only $Q(z_1, \ldots, z_N)$ exists.
\end{enumerate}
The first step can be useful on its own. 
The representation of $f$ by $y_{11}$, $\ldots$, $y_{NM}$ 
gave the lower bounds of \cite{AS}.
The new representation by $z_1$, $\ldots$, $z_N$ might yield 
new lower bounds that are easier to prove using this approach.

\subsection{New polynomial representation}

We introduce variables $z_1$, $\ldots$, $z_M$, with
$z_j=|f^{-1}(j)|$ (equivalently, $z_j=|\{i:y_{ij}=1\}|$).
We say that a polynomial $Q$ in $z_1, \ldots, z_M$
approximates $\phi$ if it satisfies requirements
similar to Definition \ref{def1}.
($Q\in[1-\epsilon, 1]$ if $\phi(f)=1$, $Q\in[0, \epsilon]$ if $\phi(f)=0$,
and $Q\in[0, 1]$ if $z_1, \ldots, z_M$ correspond to $f\in {\cal F}(N, M)$ 
for which $\phi(f)$ is not defined.)

{\bf Example 1:}
A polynomial $Q(z_1, \ldots, z_M)$ approximates 
the collision property if:
\begin{enumerate}
\item
$Q(z_1, \ldots, z_M)\in [1-\epsilon, 1]$ if $N$ of
the variables $z_1$, $\ldots$, $z_M$ are 1 and the remaining $M-N$
variables are 0;
\item
$Q(z_1, \ldots, z_M)\in [0, \epsilon]$ if $\frac{N}{2}$ of
the variables $z_1$, $\ldots$, $z_M$ are 2 and the remaining $M-\frac{N}{2}$
variables are 0;
\item
$Q(z_1, \ldots, z_M)\in [0, 1]$ if $z_1, \ldots, z_M$ are non-negative integers
and $z_1+\cdots+z_M=N$.
\end{enumerate}

{\bf Example 2:}
A polynomial $Q(z_1, \ldots, z_M)$ approximates 
element distinctness if:
\begin{enumerate}
\item
$Q(z_1, \ldots, z_M)\in [1-\epsilon, 1]$ if $N$ of
the variables $z_1$, $\ldots$, $z_M$ are 1 and the remaining $M-N$
variables are 0;
\item
$Q(z_1, \ldots, z_M)\in [0, \epsilon]$ if $z_1, \ldots, z_M$ are non-negative integers,
$z_1+\cdots+z_M=N$, and $z_i>1$ for some $i$.
\end{enumerate}

In both cases, there is no restriction on $Q(z_1, \ldots, z_M)$ when $z_1+\cdots+z_M\neq N$
because such $z_1$, $\ldots$, $z_M$ do not correspond to any $f:[N]\rightarrow [M]$.

\begin{Lemma}
\label{lem:symm}
Let $\phi:{\cal F}'\rightarrow \{0,1\}$, 
${\cal F}'\subseteq {\cal F}(N, M)$ be symmetric.
Then, the following two statements are equivalent:
\begin{enumerate}
\item[(1)]
There exists a polynomial $Q$ of degree at most $k$ 
in $z_1, \ldots, z_M$ approximating $\phi$;
\item[(2)]
There exists a polynomial $P$ of degree at most $k$ 
in $y_{11}$, $\ldots$, $y_{NM}$ approximating $\phi$.
\end{enumerate}
\end{Lemma}

\proof
To see that (1) implies (2),
we substitute $z_j=y_{1j}+y_{2j}+\ldots+y_{Nj}$ into
$Q$ and obtain a polynomial in $y_{ij}$ with the
same approximation properties.
Next, we show that (2) implies (1).

Let $P(y_{11}, \ldots, y_{NM})$ be a polynomial approximating $\phi$.
We define $Q(z_1, \ldots, z_M)$ as follows.
Let $S$ be the set of all $y=(y_{11}, \ldots, y_{NM})$
corresponding to functions $f:[N]\rightarrow [M]$
with the property that, for every $i\in[M]$ the number of
$j$ with $f(j)=i$ is exactly $z_i$.
We define $Q(z_1, \ldots, z_M)$ as the expectation of
of $P(y_{11}, \ldots, y_{NM})$ when $y=(y_{11}, \ldots, y_{NM})$
is picked uniformly at random from $S$.
(An equivalent way to define $Q$ is to fix one function $f$
with this property and to define $Q$ as the expectation of
of $P(y_{11}, \ldots, y_{NM})$, for $y=(y_{11}, \ldots, y_{NM})$
corresponding to the function $f\pi$, with $\pi$ being a random
element of $S_N$.)

Since $\phi$ is symmetric, we have $\phi(f)=\phi(f\pi)$.
Therefore, if $P(y_{11}, \ldots, y_{NM})$ approximates
$\phi$, then $Q(z_1, \ldots, z_M)$ also approximates $\phi$.
 
It remains to prove that $Q$ is a polynomial of degree at most
$k$ in $z_1, \ldots, z_M$.
Let
\[I=y_{i_1 j_1} y_{i_2 j_2} \cdots y_{i_k j_k} \]
be a monomial of $P$. It suffices to prove that each $E[I]$ is 
a polynomial of degree at most $k$ 
because $E[P]$ is the sum of $E[I]$ over all $I$.

We can assume that $i_l$ for
$l\in\{1, \ldots, k\}$ are all distinct.
(If the monomial $I$ contains two variables $y_{ij}$ 
with the same $i$, $j$, one
of them is redundant because $y_{ij}^2=y_{ij}$.
If $I$ contains $y_{ij}$, $y_{ij'}$, $j\neq j'$,
then $y_{ij}y_{ij'}=0$ because $f(i)$ cannot be 
equal $j$ and $j'$ at the same time.
Then, $I=0$.) We have
\[ E[I]=Pr[y_{i_1 j_1}=1] \prod_{l=2}^k 
Pr [ y_{i_l j_l}=1 | y_{i_1 j_1}\cdots y_{i_{l-1} j_{l-1}}=1] .\]
There are $N$ variables $y_{i j_1}$. Out of them,
$z_{j_1}$ variables are equal to 1 and each $y_{i j_1}$ is
equally likely to be 1.
Therefore,
\[ Pr[y_{i_1 j_1}=1] = \frac{z_{j_1}}{N} \]
Furthermore, let $s_{l}$ be the number of
$l'<l$ such that $j_{l}=j_{l'}$.
Then, 
\[ Pr [ y_{i_l j_l}=1 | y_{i_1 j_1}\cdots y_{i_{l-1} j_{l-1}}=1]
= \frac{z_{j_l}-s_l}{N-l-1} \]
because, once we have set $y_{i_1 j_1}=1$, $\ldots$, $y_{i_{l-1} j_{l-1}}=1$,
we have also set all other $y_{i_1 j}$, $\ldots$, $y_{i_{l-1} j}$ to 0.
Then, we have $N-l-1$ variables $y_{i j_l}$ which are not set yet and,
out of them, $z_{j_l}-s_l$ must be 1.

Therefore, $E[I]$ is a product of $k$ terms, each of which is a linear function
of $z_1, \ldots, z_M$. This means that $E[I]$ is a polynomial
in $z_1, \ldots, z_M$ of degree $k$.
This completes the proof of the lemma.
\comment{This implies that the expected value of $P$ under the
same probability distribution is a polynomial $Q$ in
$z_1, \ldots, z_M$ of degree at most $k$ as well.

Since the problem $f$ is symmetric with respect to permutations
$\pi$ on $[N]$,
applying any such permutation on $i$-index of variables $y_{ij}$
produces a polynomial with the same approximation
properties as $P$. 
Therefore, the expectatation of $P(y_{\pi(1) 1}, \ldots, y_{\pi(n) M})$ over a random
permutation $\pi\in S_N$ must have the same approximation
properties as well.}
\qed

\subsection{Lower bound for properties with small range}

We now finish the proof of Theorem \ref{MainThm}.
Obviously, the minimum degree of a polynomial
approximating $\phi'$ is at most the minimum degree of a polynomial
approximating $\phi$ (because we can take a polynomial
approximating $\phi$ and obtain a polynomial approximating $\phi'$
by restricting it to variables $y_{ij}$, $j\in[N]$).
In the other direction, we can take a polynomial $P'$
approximating $\phi'$ and obtain a polynomial $Q'$ in $z_1, \ldots, z_N$
approximating $\phi'$ by Lemma \ref{lem:symm}.
We then construct a polynomial $Q$ in 
$z_1, \ldots, z_M$ of the same degree approximating $\phi$.
After that, using Lemma \ref{lem:symm} in the other
direction gives us a polynomial $P$ in $y_{11}, \ldots, y_{NM}$
approximating $\phi$.

It remains to construct $Q$ from $Q'$. For that, we can assume
that $Q'$ is symmetric w.r.t. permuting $z_1, \ldots, z_N$.
(Otherwise, replace $Q'$ by the expectation of 
$Q'(z_{\pi(1)}, \ldots, z_{\pi(N)})$, where $\pi$ is a
uniformly random permutation of $\{1, 2, \ldots, N\}$.)
Since $Q'$ is symmetric, it is a sum 
of elementary symmetric polynomials
\[ Q'_{c_1, \ldots, c_l} =\sum_{i_1, \ldots, i_l\in[N]} 
z_{i_1}^{c_1} z_{i_2}^{c_2}\cdots z_{i_l}^{c_l} .\]
Let $Q$ be the sum of elementary symmetric polynomials
in $z_1, \ldots, z_M$ with the same coefficients.

We claim that $Q$ approximates $\phi$. To see this, consider
an input function $f:[N]\rightarrow [M]$. 
There are at most $N$ values $j\in\{1, \ldots, M\}$ such 
that there exists $i\in\{1, \ldots, N\}$ with $f(i)=j$.
This means that, out of $M$ variables $z_1, \ldots, z_M$ 
corresponding to $f$, at most $N$ are nonzero. 

Consider a permutation $\pi\in S_M$
that maps all $i\in[M]$ with $z_i\neq 0$ to $\{1, \ldots, N\}$.
Let $f'=\pi f$. 
Since $\phi$ is symmetric, $\phi(f)=\phi(f')$.
Since $f'$ is a function from $[N]$ to $[N]$,
$Q'$ correctly approximates $\phi$ on $f'$.
Since $Q(z_1,\ldots, z_N, 0, \ldots, 0)=Q'(z_1, \ldots, z_N)$,
$Q$ also correctly approximates $\phi$ on $f'$.
Since $Q$ is symmetric w.r.t. permutations of $z_1, \ldots, z_M$,
$Q$ approximates $\phi$ on the input function $f$ as well.
This completes the proof of Theorem \ref{MainThm}.

\subsection{Lower bound on the polynomial degree of the AND--OR tree}

As a by-product, our result provides a better lower bound on the 
polynomial degree of a well-studied Boolean function.

This Boolean function is the two level AND--OR tree on $N^2$ variables.
Let $x_1, \ldots, x_{N^2}\in\{0, 1\}$ be the variables. 
We split them into $N$ groups, with the $i^{\rm th}$ group 
consisting of $x_{(i-1)N+1}$, $x_{(i-1)N+2}$, $\ldots$, $x_{iN}$.
The AND--OR function $g(x_1, \ldots, x_{N^2})$ is defined as
\[ g(x_1, \ldots, x_{N^2}) = \bigwedge_{i=1}^n 
\bigvee_{j=(i-1)N+1}^{iN} x_j .\]
A polynomial $p(x_1, \ldots, x_{N^2})$ approximates $g$ if
$0\leq p(x_1, \ldots, x_{N^2})\leq\epsilon$ whenever 
$g(x_1, \ldots, x_{N^2})=0$ and 
$1-\epsilon\leq p(x_1, \ldots, x_{N^2})\leq 1$ whenever 
$g(x_1, \ldots, x_{N^2})=1$ (similarly to Definition \ref{def1}).

It has been an open problem to determine the minimum degree of a 
polynomial approximating the two-level AND--OR tree. 
The best lower bound is $\Omega(\sqrt{N \log N})$ by Shi \cite{Shi1},
while the best upper bound is $O(N)$.
(Curiously, the quantum query complexity of this problem is known.
It is $\Theta(N)$, as shown by \cite{BCW,AmbainisAdv}.
If the polynomial degree is $o(N)$, this would be the second example
of a Boolean function with a gap between the polynomial degree and
quantum query complexity, with the first example being the iterated functions
in \cite{AmbainisPoly}.)
We show

\begin{Theorem}
Any polynomial approximating $g$ has degree $\Omega(N^{2/3})$.
\end{Theorem}

\proof
Consider the element distinctness problem for $M=N$. 
An instance of this problem, $f\in{\cal F}(N, N)$ 
can be described by $N^2$ variables $y_{11}, \ldots, y_{NN}$
(as shown in Section \ref{sub:polylb}). 

The values of the function, $f(1)$, $f(2)$, $\ldots$, $f(N)$, 
are all distinct if and only if, for each $j\in[N]$,
there exists $i\in[N]$ with $f(i)=j$. 
This, in turn, is equivalent to saying that, for each
$i\in[N]$, one of the variables $y_{1i}, y_{2i}, \ldots, y_{Ni}$
is equal to 1. 

Assume we have a polynomial $P(x_1, \ldots, x_{N^2})$ of degree $d$ 
approximating the two level AND--OR tree function $g$. 
Consider the polynomial $Q(y_{11}, \ldots, y_{NN})$ obtained
from $P$ by replacing $x_{(i-1)N+j}$ with $y_{ji}$. 
If the $N$ values $f(i)$ are all distinct, then, for each 
$j\in\{1, \ldots, N\}$, there exists $i$ such that $f(i)=j$.
Therefore, one of the variables $y_{1j}, \ldots, y_{Nj}$ is 1 and 
the OR of those variables is also 1.
This means that the AND--OR 
function $g(x_1, \ldots, x_{N^2})$ is equal to 1. 
If the values $f(i)$ are not all distinct, then there 
exists $j\in[N]$ such that there is no $i$ with $f(i)=j$.
Then, $y_{1i}, y_{2i}, \ldots, y_{Ni}$ are all 0, implying
that $g(x_1, \ldots, x_{N^2})=0$ for the corresponding
assignment $x_1$, $\ldots$, $x_{N^2}$.

This means that $Q$ approximates the element distinctness property,
in the sense of section \ref{sub:polylb}.
Since degree $\Omega(N^{2/3})$ is required to approximate element 
distinctness, $d=\Omega(N^{2/3})$.
\qed

\section{Conclusion}

We have shown that, for any symmetric property of functions $f:[N]\rightarrow [M]$, 
its polynomial degree is the same for all $M\geq N$. 
Thus, if we prove a lower bound for the degree for some large $M$,
this immediately implies the same bound for $M=N$.
Since the polynomial degree is a lower bound for quantum query complexity,
this can be used to show quantum lower bounds.
As particular cases of our result, we get that the collision 
problem has degree $\Omega(N^{1/3})$ and that the element 
distinctness problem has degree 
$\Omega(N^{2/3})$, even if $M=N$. This implies $\Omega(N^{1/3})$ and 
$\Omega(N^{2/3})$ quantum lower bounds on these problems for $M=N$.

A part of our result is a new representation for polynomials describing
properties of functions $f:[N]\rightarrow [M]$. This new description
might be useful for proving new quantum lower bounds.
We conclude with two open problems. 

\begin{enumerate}
\item
{\bf Modified element distinctness problem.}
Say we are given $f:[N]\rightarrow [N]$
and we are promised that either $f$ is one-to-one or there are $i, j, k$ such that
$f(i)=f(j)=f(k)$. We would like to know which of these two is the case.
What is the quantum query complexity of this problem?

The problem is quite similar to element distinctness in which we have to
distinguish one-to-one function from one having $f(i)=f(j)$ for some $i, j$ with $i\neq j$. 
The known $O(N^{2/3})$
quantum algorithm still applies, but the $\Omega(N^{2/3})$ quantum lower bound 
of \cite{AS} (by a reduction from the collision problem) breaks down. The best lower
bound that we can prove is $\Omega(N^{1/2})$ by a reduction from Grover's search.
Improving this bound to $\Omega(N^{2/3})$ is an open problem. 

This problem is also similar to element distinctness if we look at it 
in our new $z_1$, $\ldots$, $z_M$ representation. For element distinctness, 
a polynomial $Q$ must satisfy $Q(1, \ldots, 1)\in[1-\epsilon, 1]$ and 
$Q(z_1, \ldots, z_N)\in[0, \epsilon]$ if $z_1+\cdots+z_N=N$ and $z_i\geq 2$
for some $i$. For our new problem, 
we must have $Q(1, \ldots, 1)\in[1-\epsilon, 1]$ and 
$Q(z_1, \ldots, z_N)\in[0, \epsilon]$ if $z_1+\cdots+z_N=N$ and $z_i\geq 3$
for some $i$.
In the first case, degree $\Omega(N^{2/3})$ is needed \cite{AS}.
In the second case, no such lower bound is known. 
\item
{\bf Polynomial degree vs. quantum query complexity for symmetric properties.}
Let $\phi$ be a symmetric property of functions $f:[N]\rightarrow[M]$.
Let $\deg(\phi)$ be the minimum degree of a polynomial that $\epsilon$-approximates $f$ and
$Q_2(\phi)$ be the minimum number of queries in a quantum query algorithm
computing $\phi$ with error at most $\epsilon$.
Is it true that these two quantities are polynomially related: 
$Q_2(\phi)=O(\deg^c (\phi))$ for some constant $c$?

This open problem was first proposed by Aaronson \cite{Aaronson,AS}, regarding properties 
which are only symmetric with respect to permuting inputs to $f$: $\phi(f)=\phi(f\pi)$
for any $\pi\in S_N$. It remains open both in this case and in the case
of properties having the more general symmetry considered in this paper
($\phi(f) = \phi(\sigma f \pi)$, for all $\pi\in S_N$ and $\sigma\in S_M$).
It is known that $Q_2(\phi)=O(\deg^2 (\phi))$ if $M=2$.
\end{enumerate}


\begin{thebibliography}{99}
\bibitem{Aaronson}
S. Aaronson.
Quantum lower bound for the collision problem. 
{\em Proceedings of STOC'02}, pp. 635-642.
Also quant-ph/0111102.

\bibitem{AS}
S. Aaronson, Y. Shi. Quantum lower bounds for the collision and the 
element distinctness problems. 
{\em Journal of ACM}, 51:595-605, 2004.
Earlier versions in \cite{Aaronson} and \cite{Shi}.

\bibitem{AmbainisAdv}
A. Ambainis. Quantum lower bounds by quantum arguments.
{\em Journal of Computer and System Sciences},
64:750-767, 2002. 
Earlier versions at STOC'00 and quant-ph/0002066. 

\bibitem{AmbainisPoly}
A. Ambainis. 
Polynomial degree vs. quantum query complexity.
{\em Proceedings of FOCS'03}, pp. 230-239,
quant-ph/0305028.

\bibitem{Ambainis}
A. Ambainis. Quantum walk algorithm for element distinctness.
{\em Proceedings of FOCS'04}, pp. 22-31,
quant-ph/0311001.

\bibitem{Beals}
R. Beals, H. Buhrman, R. Cleve, M. Mosca, R. de Wolf.
Quantum lower bounds by polynomials. 
{\em Journal of ACM}, 48: 778-797, 2001.
Earlier versions at FOCS'98 and
quant-ph/9802049.

\bibitem{BCW}
H. Buhrman, R. Cleve, A. Wigderson. 
Quantum vs. classical communication and computation. 
{\em Proceedings of STOC'98}, pp. 63-68.

\bibitem{BDistinct}
H. Buhrman, C. Durr, M. Heiligman, P. H\o yer, 
F. Magniez, M. Santha, and R. de Wolf.
Quantum algorithms for element distinctness. 
{\em 16th IEEE Annual Conference on Computational 
Complexity (CCC'01)}, pp.131-137,
quant-ph/0007016.

\bibitem{BCollision}
G. Brassard, P. H\o yer and A. Tapp.
Quantum algorithm for the collision problem.
{\em SIGACT News}, 28:14-19, 1997. 
Also quant-ph/9705002.

\bibitem{Counting}
G. Brassard, P. H\o yer, A. Tapp.
Quantum counting.
{\em Proceedings of ICALP'98}, pp. 820-831,
quant-ph/9805082.

\bibitem{BS}
H. Buhrman, R. \v Spalek. 
Quantum verification of matrix products,
quant-ph/0409035.

\bibitem{BWSurvey}
H. Buhrman, R. de Wolf.
Complexity measures and decision tree complexity: a survey.
{\em Theoretical Computer Science},
288:21-43, 2002. 

\bibitem{CLR}
T. Cormen, C. Leiserson, R. Rivest, C. Stein.
{\em Introduction to Algorithms,} 2nd Edition. 
The MIT Press and McGraw-Hill Book Company, 2001

\bibitem{Grover}
L. Grover. 
A fast quantum mechanical algorithm for database search.
{\em Proceedings of STOC'96}, pp. 212-219,
quant-ph/9605043.

\bibitem{Grover97}
L. Grover.
A framework for fast quantum mechanical algorithms. 
{\em Proceedings of STOC'98}, pp. 53-62, 
quant-ph/9711043.

\bibitem{HNS}
P. Hoyer, J. Neerbek, Y. Shi.
Quantum lower bounds of ordered searching, sorting and element 
distinctness. {\em Algorithmica}, 34:429-448, 2002. 
Earlier versions at ICALP'01 and quant-ph/0102078. 

\bibitem{Kutin}
S. Kutin. Quantum lower bound for the collision problem.
{\em Theory of Computing}, 1:29-36, 2005. Also
quant-ph/0304162.

\bibitem{MSS}
F. Magniez, M. Santha, and M. Szegedy. 
Quantum algorithms for the triangle problem. 
{\em Proceedings of SODA'05}.
Also quant-ph/0310134.

\bibitem{NW}
A. Nayak, F. Wu.
The quantum query complexity of approximating the median and related
statistics. {\em Proceedings of STOC'99}, pp. 384-393, 
quant-ph/9804066.

\bibitem{NS}
N. Nisan, M. Szegedy.
On the degree of Boolean functions as real polynomials. 
{\em Computational Complexity}, 4: 301-313, 1994.
Earlier version at STOC'02.

\bibitem{Shi}
Y. Shi. 
Quantum lower bounds for the collision and the element distinctness
problems. {\em Proceedings of FOCS'02}, pp. 513-519.
quant-ph/0112086.  

\bibitem{Shi1}
Y. Shi. 
Approximating linear restrictions of Boolean functions.
Manuscript.

\end{thebibliography}
\end{document}